\documentclass[12pt]{article}

\usepackage{amsmath,rotating,graphics,multicol,color} 

\newcommand{\mycaption}[2][kurz]{{\begin{center} 
\parbox{15cm}{{\bf \caption[#1]{\rm {#2}}}} \end{center} }}

\newcommand{\la}{{\mathcal L}}

\newcommand{\rxi}{$R_{\xi}$}

\newcommand{\equ}[1]{{(\ref{#1})}}

\newcommand{\sect}[1]{{Section \ref{#1}}}

\newcommand{\lr}[1]{{ \left( \, #1 \, \right) }}
\newcommand{\lreckig}[1]{{ \left[ \, #1 \, \right] }}
\newcommand{\mr}[1]{{\mathrm{#1}}}

\newcommand{\Psibar}{\overline{\Psi}}
\newcommand{\ov}[1]{{\overline{#1}}}

\begin{document}

\begin{flushright}
MC-TH-2002-01\\
WUE-ITP-2002-001\\
hep-ph/0203032
\end{flushright}

\begin{center}
{\large {\bf  Electroweak Constraints on Minimal Higher-Dimensional}}\\[0.2cm]
{\large  {\bf Extensions of the Standard  Model}}\\[0.5cm]
Alexander M\"uck$^{\,  a}$, Apostolos Pilaftsis$^{\, a,b}$ and
Reinhold R\"uckl$^{\,a}$\footnote[1]{Seminar given at the Corfu Summer
School on High Energy Physics in Corfu, Greece, September 2001.}\\[0.2cm]    
$^a${\em Institut f\"ur Theoretische Physik und Astrophysik, 
Universit\"at  W\"urzburg,\\ 
Am Hubland,  97074 W\"urzburg,  Germany}\\[0.1cm] 
$^b${\em  Department of
Physics and Astronomy, University of Manchester,\\ 
Manchester M13 9PL,
United Kingdom}
\end{center}

\begin{center} 
\parbox{14cm}{
\vspace{0.2cm} \small \centerline{\bf  ABSTRACT} 
We derive  electroweak  constraints on the   compactification scale of
minimal 5-dimensional extensions of  the Standard Model, in which  all
or  only  some of the SU(2)$_L$   and U(1)$_Y$ gauge  fields and Higgs
bosons feel the  presence of the fifth dimension.  In our analysis, we
assume that the  fermions are always localized on  a 3-brane.  In this
context, we also present  the consistent quantization procedure of the
higher-dimensional  models in the generalized  $R_\xi$ gauge.  We find
that   the  usually derived   lower bound  of    $\sim 4$~TeV  on  the
compactification scale may be significantly lowered to $\sim 3$~TeV if
the SU(2)$_L$ gauge boson is the only  particle that propagates in all
5 dimensions. }
\end{center}

\setcounter{equation}{0}
\section{Introduction}

\indent

In  the  original  formulations  of  string theory~\cite{review},  the
compactification radius  $R$ of  the  extra dimensions and the  string
mass $M_s$ were considered to be  set by the 4-dimensional Planck mass
$M_{\rm P} = 1.9\times 10^{16}$~TeV.    However, recent studies   have
shown~\cite{IA,JL,EW,ADS,DDG} that  conceivable  scenarios of  stringy
nature  may exist for  which $R$ and   $M_s$ practically decouple from
$M_{\rm P}$.  For example, in the model  of Ref.~\cite{ADS}, $M_s$ may
become as low as of  order TeV. In  this  case, $M_s$ constitutes  the
only fundamental scale in nature at which all forces including gravity
unify.  This low string-scale effective model could be embedded within
e.g.\ type I string theories~\cite{EW}, where  the Standard Model (SM)
may be  described  as   an  intersection of   higher-dimensional  $Dp$
branes~\cite{ADS,DDG,AB}.

As such intersections  may be higher dimensional as  well, in addition
to gravitons  the SM gauge  fields could also  propagate independently
within a higher-dimensional subspace  with compact dimensions of order
TeV$^{-1}$ for  phenomenological reasons. Since  such low string-scale
constructions may result in different higher-dimensional extensions of
the   SM~\cite{AB,AKT},  the   actual  experimental   limits   on  the
compactification  radius   are,  to  some   extent,  model  dependent.
Nevertheless,  most  of the  derived  phenomenological  limits in  the
literature  were  obtained  by  assuming  that  the  SM  gauge  fields
propagate    all    freely     in    a    common    higher-dimensional
space~\cite{NY,WJM,CCDG,DPQ2,RW,DPQ1,CL}.

Here,   we wish  to  lift  the  above restriction   and  focus on  the
phenomenological consequences  of  models which  minimally depart from
the  assumption of a universal higher-dimensional scenario~\cite{MPR}.
Specifically, we  will  consider 5-dimensional  extensions   of the SM
compactified  on an  $S^1/Z_2$  orbifold,   where the  SU(2)$_L$   and
U(1)$_Y$ gauge bosons may not both live in the same higher-dimensional
space, the  so-called bulk.  In  all  our models, the SM  fermions are
localized on the  4-dimensional subspace, i.e.~on  a 3-brane or, as it
is   often called,  brane.    For each  higher-dimensional   model, we
calculate the   effects  of the  fifth  dimension  on  the electroweak
observables  and     analyze  their    impact  on    constraining  the
compactification scale.

The organization of this note is as follows: in Section 2 we introduce
the basic concepts of  higher  dimensional theories in simple  Abelian
models.  After  compactifying the  extra dimensions on   $S^1/Z_2$, we
obtain  an  effective 4-dimensional  theory,  which in addition to the
usual SM states contains infinite towers of massive Kaluza--Klein (KK)
states of the  higher-dimensional   gauge fields.  In  particular,  we
consider  the    question    how    to  consistently    quantize   the
higher-dimensional models under study  in the so-called $R_\xi$ gauge.
Such a quantization procedure can  be successfully applied to theories
that include both Higgs bosons living in the bulk and/or on the brane.
After briefly discussing how  these concepts can be  applied to the SM
in Section 3, we turn our attention to the phenomenological aspects of
the models of our interest in Section 4. Because of the limited space,
technical details  are omitted in this  note.  A  complete discussion,
along with detailed  analytic results and  references, is given in our
paper  in~\cite{MPR}.  Section 5  summarizes our numerical results and
presents our conclusions.

\setcounter{equation}{0}
\section{5-Dimensional Abelian Models}\label{5DQED}

As a starting  point, let us consider the  Lagrangian of 5-dimensional
Quantum Electrodynamics (5D-QED) compactified on an $S^1/Z_2$ orbifold
given by
\begin{equation}
\label{freelagrangian}
\la (x, y) \, = \, - \frac{1}{4} F_{M N} (x, y) F^{M N} (x, y) \:
+\: \la_{\mr{GF}}(x,y)\:,
\end{equation}
where
\begin{equation}
\label{fieldstrength}
F_{M N} (x, y) \, = \, \partial_M A_N (x, y) - \partial_N A_M (x, y)
\end{equation}
denotes    the    5-dimensional    field    strength    tensor,    and
$\la_{\mr{GF}}(x,y)$ is the gauge-fixing term. The Faddeev-Popov ghost
terms have  been neglected, because the ghosts  are non-interacting in
the Abelian case.  Our notation for the Lorentz indices and space-time
coordinates  is:  $M,N  =   0,1,2,3,5$;  $\mu,\nu  =  0,1,2,3$;  $x  =
(x^0,\vec{x})$; and $y = x^5$.

In a 5-dimensional theory, the gauge-boson field $A_M$ transforms as a
vector  under  the Lorentz  group  SO(1,4).   In  the absence  of  the
gauge-fixing and ghost terms, the 5D-QED Lagrangian is invariant under
a U(1) gauge transformation:
\begin{equation}
\label{gaugetrf}
A_M(x,y) \to A_M(x,y) + \partial_M \Theta(x,y) \, .
\end{equation}
To compactify the theory on an $S^1/Z_2$ orbifold and not to spoil the
above property of gauge symmetry,  we demand for the fields to satisfy
the following equalities:
\begin{equation}
\label{fieldconstraints}
\begin{split}
A_M(x,y) \, & = \, A_M(x,y + 2 \pi R)\,, \\ 
A_{\mu}(x,y) \, & = \, A_{\mu}(x, - y)\,, \\ 
A_5(x,y) \, & = \, - A_5(x, - y)\,,\\ 
\Theta(x,y) \, & = \, \Theta(x,y + 2 \pi R)\,,\\ 
\Theta(x,y) \, & = \, \Theta(x, - y)\, .
\end{split}
\end{equation}
The field  $A_\mu (x,y)$  is taken to  be even  under $Z_2$, so  as to
embed conventional QED with a massless photon into our 5D-QED.  Notice
that  all other  constraints on  the  field $A_5(x,y)$  and the  gauge
parameter   $\Theta    (x,y)$   in   (\ref{fieldconstraints})   follow
automatically  if  the  theory  is  to remain  gauge  invariant  after
compactification.

Given the periodicity and  reflection properties of $A_M$ and $\Theta$
under $y$ in~(\ref{fieldconstraints}),  we can expand these quantities
in Fourier series, e.g.
\begin{equation}
\label{fourierseries}
A^{\mu}(x, y) \,   = \, \frac{1}{\sqrt{2 \pi R}} \, A^{\mu}_{(0)}(x)
                     \, + \, \sum_{n=1}^{\infty} \, \frac{1}{\sqrt{\pi
                     R}} \, A^{\mu}_{(n)}(x) \, \cos \lr{ \frac{n
                     y}{R}} \, , 
\end{equation}
where the  Fourier coefficients $A_{(n)}^{\mu} (x)$  are the so-called
KK  modes.  Integrating out  the $y$~dimension  we finally  obtain the
effective 4-dimensional Lagrangian
\begin{eqnarray}
\label{labeforegf}
\la(x) &=& -\, \frac{1}{4} F_{(0) \mu \nu} \, F_{(0)}^{\mu \nu} + 
   \sum_{n = 1}^{\infty} \, \Big[\, -\frac{1}{4} \, F_{(n) \mu \nu} \,
   F_{(n)}^{\mu \nu}\nonumber\\
&&+\, \frac{1}{2} \lr{ \frac{n}{R} \, A_{(n) \mu} +
   \partial_{\mu} A_{(n) 5} }\lr{ \frac{n}{R} \, A^\mu_{(n)} +
   \partial^\mu A_{(n)5} }\, \Big]\: +\:  \la_{\mr{GF}}(x) \, ,
\end{eqnarray}
where $\la_{\mr{GF}}(x) = \int_0^{2  \pi R} dy \, \la_{\mr{GF}}(x,y)$.

In  addition to  the  usual  QED terms  involving  the massless  field
$A^{\mu}_{(0)}$,  the  other terms  describe  two  infinite towers  of
massive vector  excitations $A^{\mu}_{(n)}$ and  (pseudo)-scalar modes
$A^{5}_{(n)}$ that  mix with  each other, for  $n \ge 1$.   The scalar
modes $A^{5}_{(n)}$ play the r\^ole of the would-be Goldstone modes in
a  non-linear realization  of an  Abelian  Higgs model,  in which  the
corresponding Higgs fields are taken to be infinitely massive.

The  above observation motivates us to  seek for  a higher-dimensional
generalization  of 't-Hooft's  gauge-fixing condition,  for  which the
mixing  terms  bilinear   in  $A^{\mu}_{(n)}$  and  $A^{5}_{(n)}$  are
eliminated        from        the       effective        4-dimensional
Lagrangian~\equ{labeforegf}.   Taking  advantage   of  the  fact  that
orbifold      compactification      generally      breaks      SO(1,4)
invariance~\cite{GGH}, one  can abandon the  requirement of covariance
of the gauge fixing condition  with respect to the extra dimension and
choose    the     following    non-covariant    generalized    $R_\xi$
gauge:\footnote[1]{For a recently related suggestion, see~\cite{GNN}.}
\begin{equation}
\label{gengaugefixterm}
\la_{\mr{GF}}(x, y)\ =\ -\, \frac{1}{2 \xi} (\partial^{\mu} A_{\mu}
\: - \: \xi \, \partial_5 A_5)^2 \, .
\end{equation}
Nevertheless, the gauge-fixing  term in \equ{gengaugefixterm} is still
invariant under ordinary 4-dimen\-sional Lorentz transformations. Upon
integration   over  the   extra   dimension,  all   mixing  terms   in
\equ{labeforegf} drop  out up to irrelevant total  derivatives and the
propagators for  the fields $A^{\mu}_{(n)}$ and  $A^{5}_{(n)}$ take on
their  usual  forms  that  describe  massive gauge  fields  and  their
respective  would-be  Goldstone bosons  of  an ordinary  4-dimensional
Abelian-Higgs model in the $R_\xi$ gauge:
\begin{equation}
\includegraphics{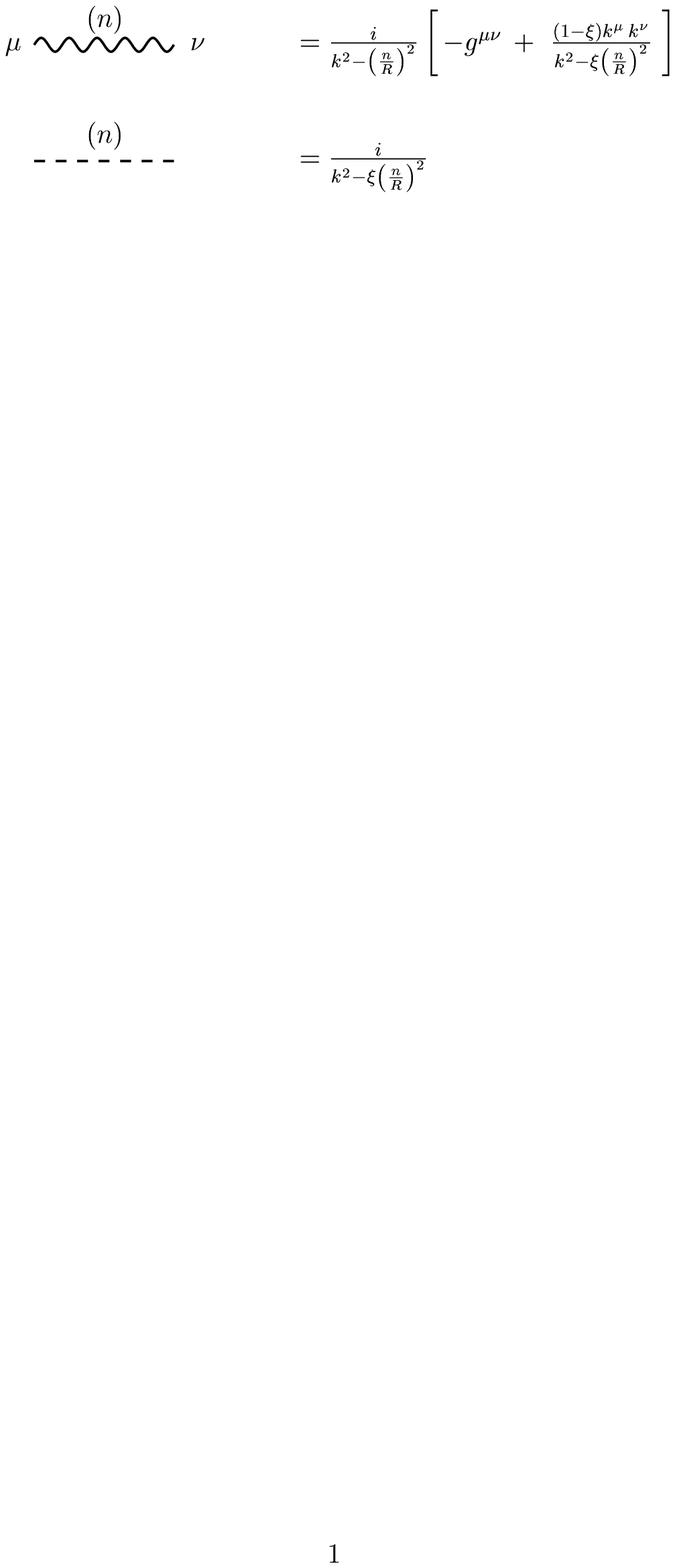}
\end{equation}
Hereafter,  we shall refer  to the  $A^{5}_{(n)}$ fields  as Goldstone
modes.

Having defined the appropriate  $R_\xi$ gauge through the gauge-fixing
term in~\equ{gengaugefixterm}, we can  recover the usual unitary gauge
in the  limit $\xi \to \infty$  \cite{PS,DMN}.  Thus, for  the case at
hand,   we    have   seen   how   starting    from   a   non-covariant
higher-dimensional gauge-fixing condition, we  can arrive at the known
covariant 4-dimensional $R_\xi$ gauge after compactification.

The above quantization procedure can now be extended to more elaborate
higher-dimen\-sional models. Adding a Higgs scalar in the bulk, the 5D
Lagrangian of the theory reads
\begin{equation}
\label{lagrdens5Dabelianhiggsmodel}
\la (x,y) \ = \ - \, \frac{1}{4} \, F^{M N}\, F_{M N}\: +\:
               (D_M\Phi)^*\, (D^M \Phi ) \: - \: V(\Phi )
               \: + \: \la_{\mr{GF}}(x,y) \, ,
\end{equation}
where $D_M = \partial_M \, + \, i \, e_5 \, A_M$ denotes the covariant
derivative, $e_5$  the 5-dimensional gauge coupling, $\Phi  (x,y)\ = \
(\, h(x,y)  \, +  \, i  \, \chi(x,y)\, )  / \sqrt{2}$  a 5-dimensional
complex scalar field, and $V(\Phi) \, =  \, \mu_5^2 \, | \Phi |^2 \, +
\, \lambda_5 \, | \Phi |^4\,$ (with $\lambda_5 > 0$) the 5-dimensional
Higgs potential.

We  consider  $\Phi   (x,y)$  to  be  even  under   $Z_2$,  perform  a
corresponding  Fourier  decomposition,  and  integrate over  $y$.  For
$\mu^2_5 < 0$,  as in the usual 4-dimensional case,  the zero KK Higgs
mode  acquires a  non-vanishing vacuum  expectation value  (VEV) which
breaks the U(1)  symmetry.  Moreover, it can be shown  that as long as
the phenomenologically relevant condition  $v < 1/R$ is met, $h_{(0)}$
will  be the  only mode  to receive  a non-zero  VEV  $\langle h_{(0)}
\rangle = v = \sqrt{2 \pi R \, |\mu_5|^2/ \lambda_5}$.

After spontaneous  symmetry breaking,  it is instructive  to introduce
the fields
\begin{equation}
\label{physunphysfieldshiggsinbulk}
G_{(n)} \, = \, \Big(\, \frac{n^2}{R^2}\: +\: e^2v^2\,
\Big)^{-1/2}\, \lr{ \frac{n}{R} \, A_{(n) 5} \: 
+\: ev\,\chi_{(n)} } \, , 
\end{equation}
where  $e  =  e_5  /  \sqrt{2  \pi  R}$,  and  the  orthogonal  linear
combinations  $a_{(n)}$. In  the effective  kinetic Lagrangian  of the
theory for the $n$-KK mode  ($n>0$), $G_{(n)}$ now plays the r\^ole of
a Goldstone  mode in  an Abelian Higgs  model, while  the pseudoscalar
field  $a_{(n)}$  describes   an  additional  physical  KK  excitation
degenerate in mass with the  KK gauge mode~$A_{(n) \mu}$ ($m^2_{a (n)}
= m^2_{A (n)} = \nolinebreak (n^2/R^2) + e^2v^2$). The spectrum of the
zero KK  modes is simply identical  to that of  a conventional Abelian
Higgs  model.   It becomes  clear  that  the appropriate  gauge-fixing
Lagrangian  in \equ{lagrdens5Dabelianhiggsmodel}  for  a 5-dimensional
generalized \rxi-gauge should be
\begin{equation}
\label{gaugefixingfunctionAbelianmodel}
\la_{\mr{GF}} (x,y) \, = \, - \, \frac{1}{2 \xi} \, \bigg[\,\partial_{\mu}
                         A^{\mu} \, -  \, \xi \, \bigg(\,\partial_5 A_5
           \: + \: e_5 \frac{v}{\sqrt{2 \pi R}} \, \chi\,\bigg)\,\bigg]^2 \, .
\end{equation}
All the mixing  terms are removed and we again  arrive at the standard
kinetic  Lagrangian for  massive  gauge bosons  and the  corresponding
would-be Goldstone  modes. The CP-odd  scalar modes $a_{(n)}$  and the
Higgs  KK-modes $h_{(n)}$  with mass  $m_{h (n)}  =  \sqrt{(n^2/R^2) +
\lambda_5  v^2  /  \pi  R}$  are  not affected  by  the  gauge  fixing
procedure.   Observe  finally   that  the   limit  $\xi   \to  \infty$
consistently corresponds to the unitary gauge.

A qualitatively  different way of  implementing the Higgs sector  in a
higher-dimensional Abelian model is to localize the Higgs field at the
$y=0$   boundary  of   the   $S^1/Z_2$  orbifold   by  introducing   a
$\delta$-function in the 5-dimensional Lagrangian
\begin{equation}
\label{lagrdens5Dabelianhiggsmodelbrane}
\la (x,y)\ =\ - \, \frac{1}{4} \, F^{M N}\, F_{M N}\: +\: 
\delta (y) \, \lreckig{ (D_{\mu} \Phi)^* \,(D^\mu \Phi)  \: 
                  - \: V(\Phi )}\: +\:  \la_{\mr{GF}}(x,y) \, ,
\end{equation}
where  the covariant  derivative and  the Higgs  potential  have their
familiar  4-dimensional   forms.   Because  the   Higgs  potential  is
effectively  four   dimensional  the  Higgs  field,   not   having  KK
excitations as a brane field, acquires the usual VEV.  Notice that the
bulk  scalar field $A_5  (x,y)$ vanishes  on the  brane $y  = 0$  as a
result  of its  odd  $Z_2$-parity and  does  not couple  to the  Higgs
sector.

After  compactification   and  integration  over   the  $y$-dimension,
spontaneous symmetry  breaking again generates  masses for all  the KK
gauge modes $A^{\mu}_{(n)}$. However,  the Fourier modes are no longer
mass  eigenstates. By diagonalization of  the mass
matrix the mass  eigenvalues $m_{(n)}$ of the KK  mass eigenstates are
found to obey the transcendental equation
\begin{equation}
\label{transcendentalequformasses}
m_{(n)} \, = \, \pi \, m^2 \, R \, \cot \lr{\pi
\, m_{(n)} \, R} \, 
\end{equation}
with  $m  = ev$. Hence,   the zero-mode mass eigenvalues  are slightly
shifted  from what we  expect in a 4D   model.  The respective KK mass
eigenstates can also be calculated analytically~\cite{MPR}.

To   find   the   appropriate    form   of   the   gauge-fixing   term
$\la_{\mr{GF}}(x,y)$   in~\equ{lagrdens5Dabelianhiggsmodelbrane},   we
follow \equ{gaugefixingfunctionAbelianmodel},  but restrict the scalar
field $\chi$ to the brane $y = 0$, viz.
\begin{equation}
\label{gaugefixingfunctionabelianhiggsonbrane}
\la_{\mr{GF}} (x,y) \, = \, - \, \frac{1}{2 \xi} \, 
        \Big[\, \partial_{\mu} A^{\mu} \ - \ 
\xi \, \big( \partial_5 \, A_5 \: + \: 
e_5 v \, \chi \, \delta(y)\big)\,\Big]^2 \, .
\end{equation}
As is expected  from a generalized $R_\xi$ gauge,  all mixing terms of
the gauge modes $A_{(n)}^{\mu}$  with $A_{(n) 5}$ and $\chi$ disappear
up to  total derivatives if $\delta (0)$  is appropriately interpreted
on  $S^1/Z_2$.   Determining  the  unphysical  mass  spectrum  of  the
Goldstone modes, we find  a one-to-one correspondence of each physical
vector mode  of mass  $m_{(n)}$ to an  unphysical Goldstone  mode with
gauge-dependent mass  $\sqrt{\xi} \,  m_{(n)}$.  In the  unitary gauge
$\xi  \to\infty$, the  would-be Goldstone  modes are  absent  from the
theory.  The  present brane-Higgs  model  does  not  predict other  KK
massive scalars apart from the physical Higgs boson $h$.

\setcounter{equation}{0}
\section{5-Dimensional Extensions of the Standard Model}
\label{MinimalExtensions}

It is a straightforward exercise to generalize the ideas introduced in
\sect{5DQED} for  non-Abelian theories.  Compactification, spontaneous
symmetry breaking and  gauge fixing are very analogous  to the Abelian
case   and   the   non-decoupling   ghost   sector   can   be   easily
included~\cite{MPR}.  Hence, in the  effective 4D theory, we arrive at
a particle spectrum  being similar to the Abelian  case.  In addition,
the self-interaction of gauge-bosons  in non-Abelian theories leads to
self-interactions of  the KK modes  which are restricted  by selection
rules reflecting the $S^1/Z_2$ structure of the extra dimension.

Turning our attention to the electroweak sector of the Standard Model,
its  gauge  structure  SU(2)$_L \otimes   $U(1)$_Y$  opens up  several
possibilities for 5-dimensional  extensions, because the SU(2)$_L$ and
U(1)$_Y$  gauge fields do not  necessarily both propagate in the extra
dimension.   Such a  realization may  be   encountered within specific
stringy  frameworks,  where one of  the  gauge  groups is  effectively
confined on the boundaries of the $S^1/Z_2$ orbifold~\cite{AB,AKT}.

However, in  the most frequently investigated  scenario, SU(2)$_L$ and
U(1)$_Y$  gauge  fields  live  in  the bulk  of  the  extra  dimension
(bulk-bulk  model).    In  this  case,   as  has  been   presented  in
\sect{5DQED},  both a  localized  (brane) and  a 5-dimensional  (bulk)
Higgs doublet can be included  in the theory.  For generality, we will
consider a 2-doublet  Higgs model, where the one  Higgs field $\Phi_1$
propagates in  the fifth  dimension, while the  other one  $\Phi_2$ is
localized.   The phenomenology of  electroweak precision  variables is
not  sensitive to details  of the  Higgs potential  but only  to their
vacuum expectation  values $v_1$ and  $v_2$, or equivalently  to $\tan
\beta = v_2 / v_1$ and $v^2 = v^2_1 + v^2_2$.

An even  more minimal  5-dimensional extension of  electroweak physics
constitutes a model in which only the SU(2)$_L$-sector feels the extra
dimension  while  the  U(1)$_Y$  gauge  field is  localized  at  $y=0$
(bulk-brane model). In  this case, the Higgs field  being charged with
respect to both gauge groups has  to be localized at $y=0$ in order to
preserve gauge invariance of  the (classical) Lagrangian. For the same
reason, a bulk Higgs is forbidden in the third possible model in which
SU(2)$_L$  is  localized  while   U(1)$_Y$  propagates  in  the  fifth
dimension (brane-bulk model).

In all these minimal 5-dimensional extensions of the SM we assume that
the  SM  fermions are  localized  at the  $y=0$   fixed  point of  the
$S^1/Z_2$ orbifold.  The  coupling of such  a fermion to a gauge boson
restricted to the same  brane $y=0$ has its SM   value.  On the  other
hand, the effective interaction Lagrangian  describing the coupling of
a fermion to the  Fourier modes of a  bulk gauge-boson has the generic
form
\begin{equation}
\label{interactionsfermwithgbmodesabelianmodel}
\la_{\rm int}(x)\ = \ \Psibar\, \gamma^{\mu} \, \lr{g_V
                        + g_A \gamma^5} \, \Psi\, \Big( A_{(0)
                        \mu} \, + \, \sqrt{2} \, \sum_{n=1}^{\infty}
                        A_{(n) \mu} \Big) \, .
\end{equation}
Again, the coupling parameters $g_V$  and $g_A$ are set by the quantum
numbers of the  fermions and receive their SM  values.  Because the KK
mass  eigenmodes  generally  differ  from  the  Fourier  modes,  their
couplings to fermions $g_{V(n)}$  and $g_{A(n)}$ have to be calculated
for   each   model   individually,   after   the   appropriate   basis
transformations relating  the weak  to mass eigenstates  have properly
been taken into account.

\setcounter{equation}{0}
\section{Effects on Electroweak Observables}
\label{Phenomenology}

In this section, we will  concentrate on the phenomenology and present
bounds   on  the  compactification   scale  $M   =  1/R$   of  minimal
higher-dimensional  extensions of  the  SM calculated  by analyzing  a
large  number  of  high  precision  electroweak  observables.   To  be
specific,  we  proceed  as  follows.   We  relate  the  SM  prediction
$\mathcal{O}^{\mr{SM}}$~\cite{PDG}  for an  electroweak  observable to
the  prediction  $\mathcal{O}^{\mr{HDSM}}$  for  the  same  observable
obtained in the higher-dimensional SM under investigation through
\begin{equation}
\label{generalformofpredictions}
{\cal O}^{\rm HDSM} \ =\ {\cal O}^{\rm SM} \, \big( 1\: +\: 
\Delta^{\rm HDSM}_{\cal O} \big)\, .
\end{equation}
Here, $\Delta^{\rm HDSM}_{\cal O}$ is the tree-level modification of a
given observable ${\cal O}$  from its SM  value due to the presence of
one   extra  dimension.   In   five  dimensions,  all   the tree-level
modifications can be expanded in  powers  of the typical scale  factor
$X=\frac{1}{3}\pi^2 m^2_Z R^2$.  On the other hand, to enable a direct
comparison    of   our predictions   with   the  electroweak precision
data~\cite{PDG},   we     include      SM radiative        corrections
to~$\mathcal{O}^{\mr{SM}}$.   However,  we   neglect  SM- as  well  as
KK-loop contributions to $\Delta^{\rm  HDSM}_{\cal O}$ as higher order
effects.

As input SM parameters for  our theoretical predictions, we choose the
most accurately measured ones,  namely  the $Z$-boson mass $M_Z$,  the
electromagnetic  fine   structure  constant~$\alpha$  and    the Fermi
constant $G_F$.  While  $\alpha$ is not  affected in the models  under
study, $M_{Z}$  and $G_F$  generally deviate  from their  SM  form. To
first order in $X$, $M_{Z}$ and $G_F$ may be parameterized as
\begin{equation}
\label{deltazdef}
M_{Z} \  =\ M^{\mr{SM}}_{Z} \, \lr{1 \: + \: \Delta_{Z}\,X}\,, \qquad
G_F \ = \ G^{\mr{SM}}_F \, \lr{ 1\: +\: \Delta_G\, X}\,,
\end{equation}
where $\Delta_{Z}$ and  $\Delta_G$ are model-dependent parameters. For
example, one finds
\begin{equation}
\Delta_{Z} = \big\{ \ - \, \frac{1}{2} \,  \sin^4\beta\,,\ - 
\, \frac{1}{2} \, \sin^2 \hat{\theta}_w\,,\ - \, \frac{1}{2} \, 
\cos^2\hat{\theta}_w\, \big\}\,.
\end{equation} 
for the  bulk-bulk, brane-bulk and bulk-brane models,  with respect to
the SU(2)$_L$ and U(1)$_Y$ gauge groups.

The relation  between the weak  mixing angle $\theta_w$ and  the input
variables is also affected by the fifth dimension. Hence, it is useful
to  define  an  effective  mixing angle  $\hat{\theta}_w$  by  $\sin^2
\hat{\theta}_w  \ =  \nolinebreak \  \sin^2 \theta_w  \, \lr{  1\: +\:
\Delta_{\theta}\,X }\, ,$ such that the effective angle still fulfills
the tree-level relation
\begin{equation}
\label{definitionequationfonormalizedthetaw}
G_F \ = \ \frac{\pi \alpha}{\sqrt{2} \sin^2 \hat{\theta}_w \, \cos^2
\hat{\theta}_w \, M_{Z}^2} \ ,
\end{equation}
of the Standard Model.

For the tree-level calculation  of $\Delta^{\rm HDSM}_{\cal O}$, it is
necessary to  consider the mixing effect  of the Fourier  modes on the
masses  of  the  Standard-Model  gauge  bosons as  well  as  on  their
couplings to fermions.  In addition, we have to keep  in mind that the
mass spectrum of  the KK gauge bosons also depends  on the model under
consideration.

Within   the   framework  outlined  above,  we   compute  $\Delta^{\rm
HDSM}_{\cal O}$ for the  following high precision observables to first
order in $X$: the $W$-boson  mass $M_W$, the $Z$-boson invisible width
$\Gamma_Z     (\nu    \ov{\nu})$,     $Z$-boson     leptonic    widths
$\Gamma_Z(l^+l^-)$, the $Z$-boson hadronic width $\Gamma_Z(\mr{had})$,
the  weak charge of  cesium $Q_W$  measuring atomic  parity violation,
various  ratios $R_l$  and $R_q$  involving partial  $Z$-boson widths,
fermionic  asymmetries $A_f$ at  the $Z$  pole, and  various fermionic
forward-backward  asymmetries $A_{\mr{FB}}^{(0,f)}$. For  example, for
the invisible $Z$ width $\Gamma_Z(\nu \overline{\nu})$ we obtain
\begin{equation}
\Delta^{\rm HDSM}_{\Gamma_Z(\nu \overline{\nu})} = \left\{
\begin{array}{cl}
\sin^2 \! \hat{\theta}_w \, \lr{\! \sin^2 \! \beta - 1 \!}^2 - 1 & \mbox{for  the bulk-bulk  model,}\\[1ex]
- \sin^2 \! \hat{\theta}_w & \mbox{for the  brane-bulk model,} \\[1ex]
- \cos^2 \! \hat{\theta}_w & \mbox{for the bulk-brane model.}
\end{array}
\right.
\end{equation}

\begin{figure}[t]
\begin{center}
\includegraphics[width=7.2cm]{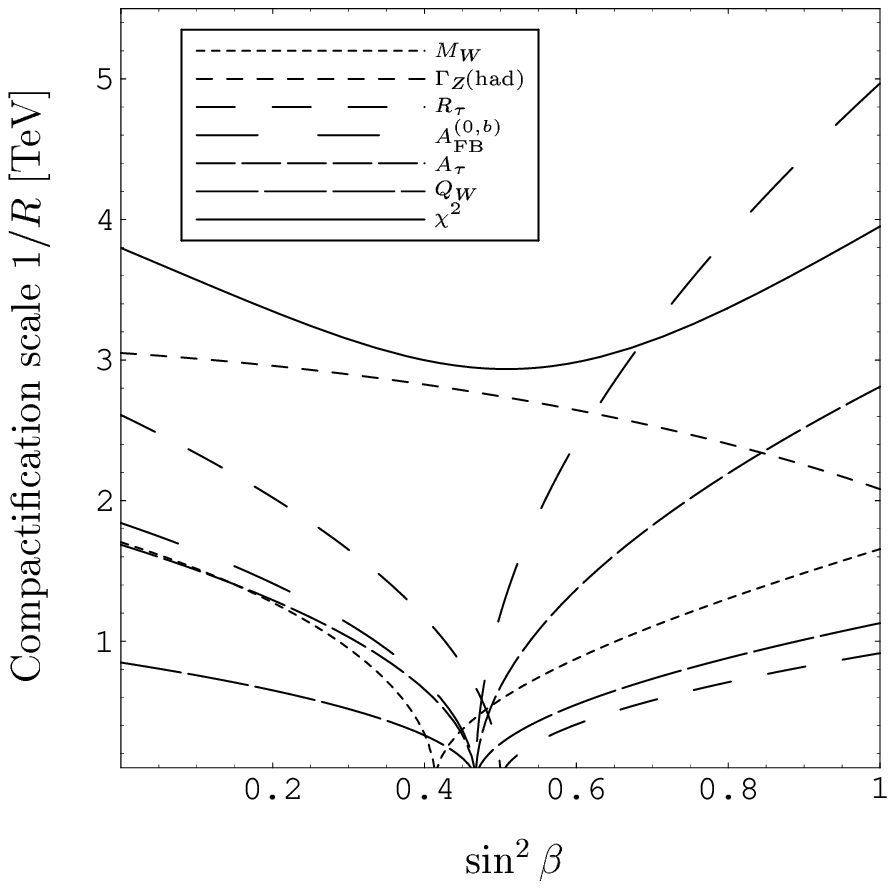}
\hspace{24pt}
\includegraphics[width=7.8cm]{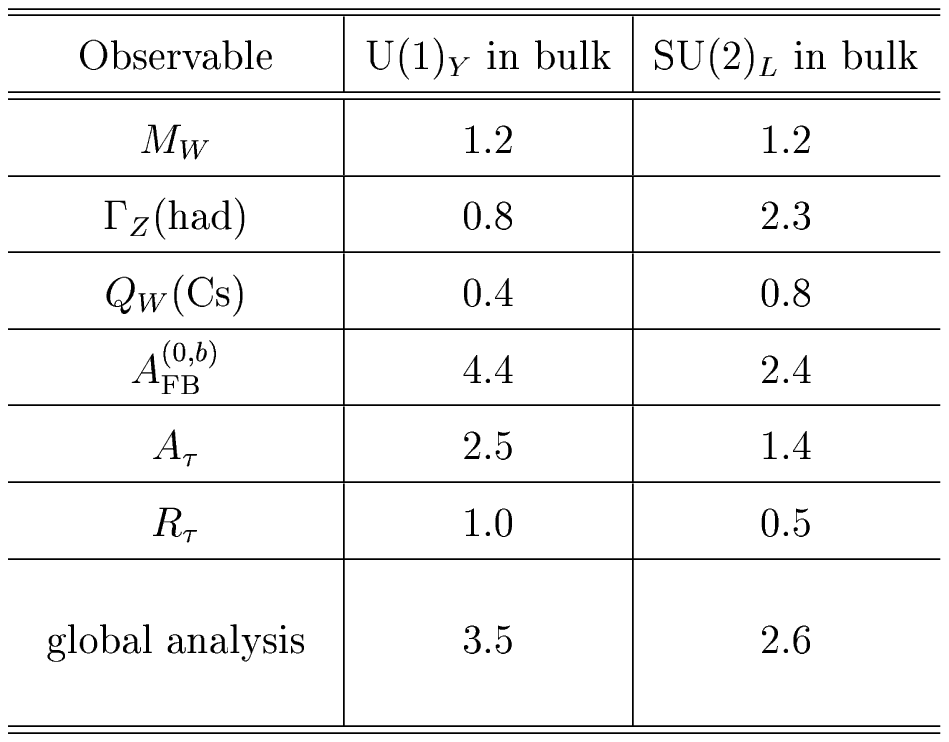}
\end{center}
\vspace{-13pt}
\hspace{2pt}
\parbox{210pt}{
{\bf
\caption{\label{bulkbulkboundlot} \rm \hfill Lower bounds on  $M=1/R$ \newline (in TeV) 
at  the 3$\sigma$ confidence level for the bulk-bulk model .}
}}
\end{figure}
\begin{table}[t]
\vspace{-60pt}
\hspace{242pt}
{\bf
\parbox{210pt}{
\caption{\rm \label{branebulkboundplot} \rm \hfill Lower bounds on  $M=1/R$ \newline (in TeV) 
at  the 3$\sigma$ confidence level for the brane-bulk and bulk-brane models.}
}}
\end{table}

Employing   the results   of    $\Delta^{\mr{HDSM}}_{\mathcal{O}}$ and
calculating all the electroweak observables considered in our analysis
by   virtue   of~\equ{generalformofpredictions},  we  confront   these
predictions with the respective   experimental values.  We can  either
test  each variable individually or perform  a $\chi^2$ test to obtain
bounds on the compactification scale $M=1/R$, where
\begin{equation}
  \label{chi2}
\chi^2(R) \ = \ \sum_i \,\frac{\lr{ \mathcal{O}_i^{\mr{exp}}\: -\:
\mathcal{O}_i^{\mr{HDSM}} }^2}{\lr{ \Delta \mathcal{O}_i}^2} \, ,
\end{equation}
$i$ runs  over all the  observables and $\Delta \mathcal{O}_i$  is the
combined experimental and theoretical error.

\begin{table}[t]
\begin{center}
\includegraphics[width=14cm]{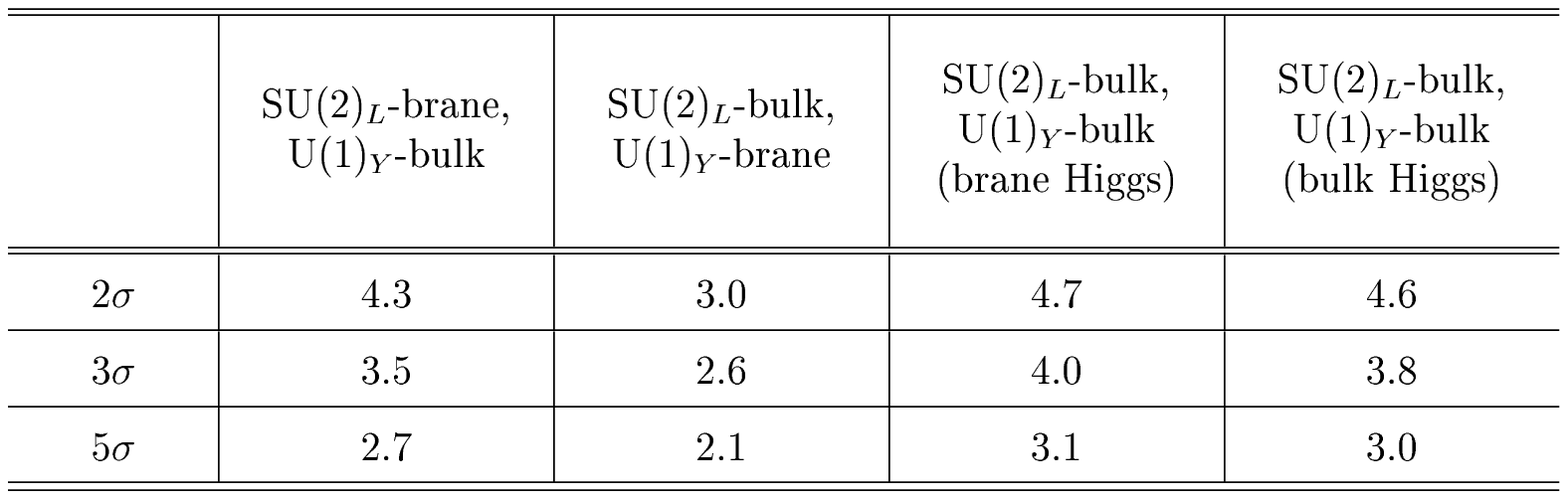}
\end{center}
\mycaption{\label{globalbounds}  Lower bounds (in TeV) on the
compactification scale $M = 1/R$ at 2$\sigma$, 3$\sigma$ and 5$\sigma$
confidence levels.}
\vspace{-0.5cm}
\end{table}

Figure~\ref{bulkbulkboundlot}  summarizes the   lower  bounds   on the
compactification scale   $M =  1/R$  coming from   different types  of
observables  for the bulk-bulk model.   In this model, we present  the
bounds as a function of $\sin^2\beta$ parameterizing the Higgs sector.
In Table~\ref{branebulkboundplot}, we summarize the bounds obtained by
our global fits  for the two  bulk-brane models. The  bounds resulting
for different confidence levels are given in Table~\ref{globalbounds}.

\setcounter{equation}{0}
\section{Discussion and Conclusions}

\indent

By performing $\chi^2$-tests, we obtain different sensitivities to the
compactification radius $R$ for  the three models under consideration:
(i)~the SU(2)$_L\otimes$U(1)$_Y$-bulk model, where all SM gauge bosons
are bulk fields; (ii)  the SU(2)$_L$-brane, U(1)$_Y$-bulk model, where
only the SU(2)$_L$  fields are restricted to the  brane, and (iii)~the
SU(2)$_L$-bulk,  U(1)$_Y$-brane model, where  only the  U(1)$_Y$ gauge
field is  confined to  the brane.  The  strongest bounds hold  for the
often-discussed bulk-bulk model no matter if the Higgs boson is living
in the  bulk or on the  brane.  For the bulk-brane  models, we observe
that the bounds  on $1/R$ are significantly reduced  by an amount even
up to  1.4~TeV for 3$\sigma$ if  the SU(2) bosons are  the only fields
that propagate in the bulk.

The lower limits on the compactification scale  derived by the present
global analysis indicate that   resonant  production of the  first  KK
state  may  be accessed at  the  LHC, at  which heavy  KK masses up to
6--7~TeV~\cite{AB,RW}   might  be  explored.   In particular,   if the
$W^\pm$ bosons propagate in the bulk with a compactification radius $R
\sim 3$~TeV$^{-1}$,  one may even  be able to probe  resonant effects
originating from the  second KK state, and so  differentiate the model
from other 4-dimensional new-physics scenaria.

In addition,  we have paid special attention  to consistently quantize
the  higher-dimen\-sional models  in the  generalized  $R_\xi$ gauges.
Specifically,  we   have  been   able  to  identify   the  appropriate
higher-dimensional gauge-fixing conditions  which should be imposed on
the theories  so as to yield  the known $R_\xi$ gauge  after the fifth
dimension  has been  integrated out.\footnote[2]{After  \cite{MPR} had
been communicated, we became aware of \cite{DCH}, which also discusses
the $R_\xi$  gauge before compactification  in fermionless non-Abelian
theories.}

\subsection*{Acknowledgements}
This  work was supported  by the  Bundesministerium f\"ur  Bildung and
Forschung (BMBF,  Bonn, Germany) under the  contract number
05HT1WWA2.

\newpage

\begin{multicols}{2}

\end{multicols}

\end{document}